\documentclass[12pt]{article}
\usepackage{mathtext}
\usepackage[utf8]{inputenc}
\usepackage[english]{babel}
\usepackage{graphicx}
\usepackage{amssymb}

\begin{document}

\title{MOLECULES IN THE EARLY UNIVERSE}
\author{B. Novosyadlyj$^1$\footnote{novos@astro.franko.lviv.ua}, O. Sergijenko$^1$\footnote{olka@astro.franko.lviv.ua}, V. M. Shulga$^{2,3}$\footnote{shulga@rian.kharkov.ua}}

\maketitle

\medskip
\centerline{\it $^1$Astronomical Observatory of 
Ivan Franko National University of Lviv,}

\centerline{\it Kyryla i Methodia str., 8, Lviv, 79005, Ukraine}
  
\centerline{\it $^2$Jilin University, Qianjin str., 2699,
Changchun, 130012, P. R. China} 

\centerline{\it $^3$Institute of Radio Astronomy NAS of Ukraine, Mystetstv str., 4, Kharkiv, 61002, Ukraine}

\begin{abstract}
We study the formation of first molecules, negative Hydrogen ions and molecular 
ions in model of the Universe with cosmological constant and cold dark matter. 
The cosmological recombination is described in the framework of modified 
model of the effective 3-level atom, while the kinetics of chemical reactions 
in the framework of the minimal model for Hydrogen, Deuterium and Helium. It is 
found that the uncertainties of molecular abundances caused by the inaccuracies
of computation of cosmological recombination are about 2-3\%.
The uncertainties of values of cosmological parameters affect the abundances of
molecules, negative Hydrogen ions and molecular ions at the level of up to 2\%. 
In the absence of cosmological reionization at redshift $z=10$ the ratios of abundances to the Hydrogen one are
$3.08\times10^{-13}$ for $H^-$, $2.37\times10^{-6}$ for $H_2$, 
$1.26\times10^{-13}$ for $H_2^+$, $1.12\times10^{-9}$ for $HD$ and 
$8.54\times10^{-14}$ for $HeH^+$.
\end{abstract}

\section*{Introduction}

At the early stages of evolution of the Universe ($z\sim10^4$) all atoms of Hydrogen, Deuterium, Helium and Lithium are fully ionized by the thermal radiation (for details see \cite{peebles1968,seager1999,seager2000}). After emergence of the neutral atoms formation of the first molecules begins. This process is widely studied (in particular, \cite{galli1998,galli2013,gloabel08,glosav08,izotov1984,lep02,pfe03,puy1993,puysig07,
sig99,sta98,vonlanthen2009}) because of its importance for cooling of the gas clouds from which  the first luminous objects have formed.

For calculation of the evolution of number densities of the first molecules it is important to know precisely the evolution of number densities of the ionized fractions during the cosmological recombination. Most often both recombination and formation of the first molecules are described by the equations of chemical kinetics with the analytical approximations for reaction rates (e. g., \cite{galli1998,vonlanthen2009}). This approach allows to do fast computations but is insufficiently accurate. Another method is based on taking into account accurately the transitions in multilevel atoms both for recombination and formation of molecules (e. g. \cite{alizadeh2011}). Its sufficient weakness is the slowness of computations. For the fastest computation of evolution of the number densities of neutral atoms and ions of Hydrogen and Helium during the cosmological recombination with the accuracy comparable to the model of multilevel atom the modified model of effective 3-level atom \cite{seager1999,seager2000} has been proposed. For study of the first molecules formation it was used e. g. in \cite{hirata2006,schleicher2008}.

To test cosmological models with the \textit{Planck} satellite data the fast computation of free electrons number density in the Universe with the uncertainty no more than few tenths of a percent is necessary. To obtain such precision the model of effective N-level atom has been developed and the codes  HyREC \cite{hyrec} and CosmoRec \cite{cosmorec} have been created. As an alternative the modified model of effective 3-level atom has been complemented \cite{ms2009,wms2008}.

The goal of this work is to investigate formation of the primordial molecules, negative Hydrogen ions and molecular ions in the framework of modified model of the effective 3-level atom for cosmological recombination and the minimal model for kinetics of chemical reactions, and to study the effect of accuracy of the cosmological recombination description and the cosmological parameter values on it. We restrict ourselves to the cosmological $\Lambda$CDM model and neglect the reionization of medium.

\section{Evolution equations for the number densities of chemical species in the Universe}

After the beginning of recombination the primordial medium consists of neutral atoms, ions and molecules of Hydrogen, Deuterium, Helium and Lithium, photons of the thermal background radiation and cold dark matter particles. We consider the last ones to be stable and interacting only via gravitation and maybe weak force, thus they have no effect on the kinetics of recombination and dissociation of atoms and molecules. Number densities of these components are determined by the characteristic for them chemical reactions, the number densities of reactants and the reaction rates depending on the temperature of matter and radiation in the medium.

The kinetics of chemical reactions is described by the equations \cite{galli1998,puy1993,vonlanthen2009}:
\begin{equation}
\left(\frac{dx_i}{dt}\right)_{\rm{chem}}=\sum_{mn}k_{mn}f_{\tilde{m}}f_{\tilde{n
}}x_{m}x_{n}+\sum_{m}k_{m\gamma}f_{\tilde{m}}x_{m}-\sum_{j}k_{ij}f_{\tilde{i}}f_
{\tilde{j}}x_{i}x_{j}-k_{i\gamma}f_{\tilde{i}}x_{i},
\label{cin}
\end{equation}
where $k_{mn}$ -- reaction rates for the reactants $m$ and $n$; 
$f_{\tilde{m}}$ is $f_{He}=n_{He}/n_H$ for reactants $m$ containing Helium, $f_D=n_D/n_H$ for reactants $m$ containing Deuterium and $f_H=n_{H}/n_H\equiv1$ for reactants $m$ containing only Hydrogen. For chemical species containing only Hydrogen the abundance is $x_m=n_m/n_H$, where $n_m$ is the number density of species $m$, $n_H$ is the total number density of Hydrogen; for species containing Deuterium, Helium and Lithium $x_m=n_m/n_D$, $x_m=n_m/n_{He}$ and $x_m=n_m/n_{Li}$, where $n_D$, $n_{He}$ and $n_{Li}$ are the total number densities of Deuterium, Helium and Lithium.

In the Universe with Friedmann-Robertson-Walker metric it is convenient to replace the time derivatives by the derivatives with respect to the redshift as:
\begin{eqnarray*}
\frac{d}{dz} = -\frac{1}{H(1+z)}\frac{d}{dt},
\label{dz}
\end{eqnarray*}
where $H\equiv(da/dt)/a$ is the Hubble parameter. 

We should note that we do not consider the energy exchange between the matter and radiation via the functions of molecular cooling and heating, because due to the small number densities of the primordial molecules in the homogeneous Universe this has no significant effect on evolution of the matter temperature \cite{hirata2006,puy1996,puy1997}.

\section{Minimal model of chemical reactions in the primordial medium}

Chemical processes in the primordial medium containing Hydrogen, Deuterium, Helium and Lithium are described by the full model \cite{galli1998} consisting of 87 reactions. However, in the study of kinetics of primordial molecules formation we use the minimal model \cite{galli1998} which is somewhat simplified but is sufficient for the correct computation of the number densities of chemical species.

The minimal model consists of 33 reactions: 10 for Hydrogen, 6 for Deuterium, 3 for Helium and 14 for Lithium. As the fractional abundance of Lithium is almost negligible ($f_{Li}\equiv 
n_{Li}/n_H\sim10^{-10}$) and there are sufficient uncertainties in its determination and the chemical network without Lithium is closed, we restrict ourselves only to consideration of Hydrogen, Deuterium and Helium.

Chemical reactions of the minimal model for $H$, $D$ and $He$ are presented in the Table \ref{reaction}. Reaction rates are taken from \cite{galli1998}, except for the rates of recombination and photoionization of Hydrogen and Helium.

\begin{table}[ht]
\caption{Chemical reactions (numbers according to \cite{galli1998})}
\label{reaction}
\centering
\begin{tabular}{l l l l}
\hline
(H1)&$\mathrm{H^+ + e^- \rightarrow H + \gamma}$&(H2)&$\mathrm{H + \gamma 
\rightarrow 
H^+ + e^-}$\\
(H3)&$\mathrm{H + e^- \rightarrow H^- + \gamma}$&(H4)&$\mathrm{H^- + \gamma 
\rightarrow H + e^-}$\\
(H5)&$\mathrm{H^- + H \rightarrow H_2 + e^-}$&(H7)&$\mathrm{H^- + H^+ 
\rightarrow H 
+H}$ \\
(H8)&$\mathrm{H + H^+ \rightarrow H_2^+ + \gamma}$&(H9)&$\mathrm{H_2^+ + \gamma 
\rightarrow H + H^+}$ \\
(H10)&$\mathrm{H_2^+ + H \rightarrow H_2 + H^+}$&(H15)&$\mathrm{H_2 + H^+ 
\rightarrow 
H_2^+ + H}$\\
\hline
(D1)&$\mathrm{D^+ + e^- \rightarrow D + \gamma}$&(D2)&$\mathrm{D + \gamma 
\rightarrow D^+ + e^-}$\\
(D3)&$\mathrm{D + H^+ \rightarrow D^+ + H}$&(D4)&$\mathrm{D^+ + H \rightarrow D 
+ 
H^+}$\\
(D8)&$\mathrm{D^+ + H_2 \rightarrow H^+ + HD}$&(D10)&$\mathrm{HD + H^+ 
\rightarrow 
H_2 + D^+}$\\
\hline
(He8)&$\mathrm{He + H^+ \rightarrow HeH^+ + \gamma}$&&\\
(He11)&$\mathrm{HeH^+ + H \rightarrow He + H_2^+}$&(He14)&$\mathrm{HeH^+ + 
\gamma 
\rightarrow He + H^+}$\\
\hline
\end{tabular}
\end{table}

\section{Cosmological recombination}

In thermodynamic equilibrium the recombination of $HeIII$, $HeII$ and $HII$ is described by the Saha equations. At $z\sim 8000$ HeIII begins to recombine. From the Saha equations for recombination of $HeIII\rightarrow HeII$ and $HeII\rightarrow HeI$ it is possible to obtain equation for the free electron fraction $x_e=n_e/n_H$ (Hydrogen and Deuterium are at this stage fully ionized):
\begin{eqnarray}
x_e^3+x_e^2\left(\eta_{HeI}-1-f_D\right)+x_e\eta_{HeI}\left(\eta_{HeII}-1-f_D-f_
{He}\right)-\eta_{HeI}\eta_{HeII}\left(1+f_D+2f_{He}\right)=0,
\end{eqnarray}
where:
\begin{eqnarray*}
&&\eta_{HeII}=\frac{(2\pi m_ekT_m)^{3/2}}{h^3n_H}e^{-\chi_{HeII}/kT_m},\\
&&\eta_{HeI}=4\frac{(2\pi m_ekT_m)^{3/2}}{h^3n_H}e^{-\chi_{HeI}/kT_m}.
\end{eqnarray*}

When the whole HeIII recombines to HeII, the Saha equations for recombination of $HeII\rightarrow HeI$ and $HII\rightarrow HI$ yield the following equation for $x_e$ (see also \cite{novos2007,novos2006}):
\begin{eqnarray}
x_e^3+x_e^2\left(\eta_{HI}+\eta_{HeI}\right)+x_e\left(\eta_{HI}\eta_{HeI}-\eta_{
HI}\left(1+f_D\right)-\eta_{HeI}f_{He}\right)-\eta_{HI}\eta_{HeI}\left(1+f_D+f_{
He}\right)=0,
\end{eqnarray}
where:
\begin{eqnarray*}
\eta_{HI}=\frac{(2\pi m_ekT_m)^{3/2}}{h^3n_H}e^{-\chi_{HI}/kT_m}.
\end{eqnarray*}
Hereafter we describe the Deuterium recombination similarly to the Hydrogen one. 

The real solutions of these cubic equations give the values of the relative number density of free electrons from which the abundances of ionized and neutral Hydrogen, Deuterium and Helium can be computed (see \cite{novos2007,novos2006}).

It should be noted that at the stage of joint equilibrium recombination of HII and HeII the first molecules, negative and molecular ions begin to form, their abundances can be computed using the formulas from appendix \ref{a}. These solutions set the initial conditions for integration of equations (\ref{cin}) at the stage of non-equilibrium recombination.

When the thermodynamic equilibrium breaks down, the modified model of effective 3-level atom \cite{seager1999} is used to describe the kinetics of non-equilibrium kinetics of recombination. The evolution of abundances of ionized Hydrogen and Helium is governed by the equations:
\begin{eqnarray}
&&\frac{d\;x_{HII}}{d\;z}=\left[x_ex_{HII}n_H\alpha_H-\beta_H\left(1-x_{HII}
\right)\exp{\left(-\frac{h\nu_{H2s}}{k_BT_m}\right)}\right]\frac{C_H}{
H\left(1+z\right)},\label{h}\\
&&\frac{d\;x_{HeII}}{d\;z}=\left[x_ex_{HeII}n_H\alpha_{He}^s-\beta_{He}
^s\left(1-x_{HeII}\right)\exp{\left(-\frac{h\nu_{He2^1s}}{k_BT_m}\right)}\right]
\frac{C_{He}^s}{H\left(1+z\right)}\nonumber\\
&&+\left[x_ex_{HeII}n_H\alpha_{He}^t-\frac{g_{He2^3s}}{g_{He2^1s}}\beta_{He}
^t\left(1-x_{HeII}\right)\exp{\left(-\frac{h\nu_{He2^3s}}{k_BT_m}\right)}\right]
\frac{C_{He}^t}{H\left(1+z\right)},\label{he}
\end{eqnarray}
where the correction factors are:
\begin{eqnarray*}
&&C_H=\frac{1+K_H\Lambda_Hn_H\left(1-x_{HII}\right)}{
1+K_H\left(\Lambda_H+\beta_H\right)n_H\left(1-x_{HII}\right)},\\
&&C_{He}^s=\frac{1+K_{He}^s\Lambda_{He}n_Hf_{He}\left(1-x_{HeII}\right)\exp{
\left(\frac{h\nu_{ps}}{k_BT_m}\right)}}{1+K_{He}^s\left(\Lambda_{He}+\beta_{He}
^s\right)n_Hf_{He}\left(1-x_{HeII}\right)\exp{\left(\frac{h\nu_{ps}}{k_BT_m}
\right)}},\\
&&C_{He}^t=\frac{1}{1+K_{He}^t\beta_{He}^tn_Hf_{He}\left(1-x_{HeII}\right)\exp{
\left(\frac{h\nu_{ps}^t}{k_BT_m}\right)}}.
\end{eqnarray*}
Here we use the notation and values of the coefficients and atomic constants from \cite{seager1999,seager2000,wms2008}.

The equation (\ref{h}) and the first term in (\ref{he}) constitute the basic modified model of effective 3-level atom proposed in \cite{seager1999,seager2000} (for a review see also\cite{novos2007}). Further studies of the cosmological recombination, caused by the need to obtain the accuracy of the $x_e$ computation sufficient for testing of the cosmological models with the \textit{Planck} satellite data, have led to the complementing of this model.

The full model of effective 3-level atom  \cite{wms2008} takes into account additionally 
the full expression for escape probability for the singlet $2^1p\rightarrow 
1^1s$ transition of $HeI$, the effect of continuum opacity of Hydrogen on $HeI$ singlet \cite{kiv2007}, the recombination through the triplet $2^3p\rightarrow 1^1s$ transition of $HeI$ (the second term in (\ref{he})) and the the effect of continuum opacity of Hydrogen on it. For Hydrogen the additional correction function is applied to the term taking into account the reddening of $Ly\alpha$-photons due to the expansion of the Universe \cite{rubinomartin2010}:
\begin{eqnarray*}
K_H=\frac{\lambda_{H2p}^3}{8\pi 
H(z)}\left[1+\exp\left(-\frac{(\log(1+z)-z_1)^2}{w_1^2}\right)+\exp\left(-\frac{
(\log(1+z)-z_2)^2}{w_2^2}\right)\right].
\end{eqnarray*}
This allows to fit the modified model of effective 3-level atom to the model of effective N-level atom.

The matter temperature $T_m$ is virtually equal to the radiation temperature $T_r$ down to $z\sim 850$. The rate of change the temperature is described by the adiabatic cooling of radiation due to the expansion of the Universe:
\begin{equation}
  \label{Tm1}
  \frac{d\;T_m}{d\;z}=\frac{T_m}{1+z}.
\end{equation}
Later, at redshifts $z\lesssim850$, the following equation for the rate of change of the matter temperature is used \cite{seager1999}:
\begin{equation}
  \label{Tm2}  
\frac{d\;T_m}{d\;z}=\frac{2T_m}{1+z}+\frac{8\sigma_Ta_rT_r^4}{3m_ecH(1+z)}\frac{
x_e}{1+f_D+f_{He}+x_e}\left(T_m-T_r\right).
\end{equation}
To smooth the transition from (\ref{Tm1}) to (\ref{Tm2}) the correction \cite{ms2009} is applied.

The system of equations (\ref{h})-(\ref{Tm2}) is complemented by the equations for Deuterium recombination and for kinetics of chemical reactions (\ref{cin}).

\section{Results and discussion}

On basis of the code for computation of evolution of the relative number density of free electrons during the cosmological recombination using the modified model of effective 3-level atom recfast\footnote{http://www.astro.ubc.ca/people/scott/recfast.html} (version 1.5.2) we developed the code for computation of the abundances of primordial negative Hydrogen ions and molecular ions $H^-$, $H_2^+$, $HeH^+$ and molecules $H_2$, $HD$. The system of equations describing simultaneously the cosmological recombination and the kinetics of chemical reactions is stiff, therefore we use the Gear method of integration implemented in the code DDRIV\footnote{http://www.netlib.org/slatec/src/ddriv1.f} which is in open access.

\begin{figure}
 \label{HDHe}
\includegraphics[width=0.32\textwidth]{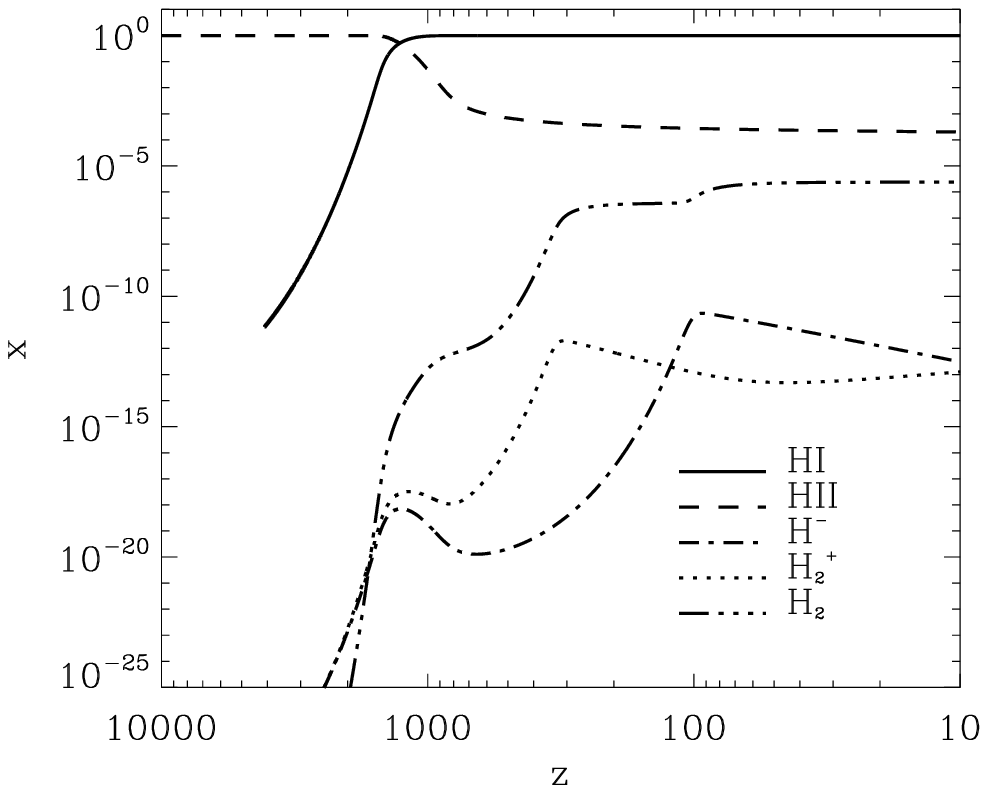}
\includegraphics[width=0.32\textwidth]{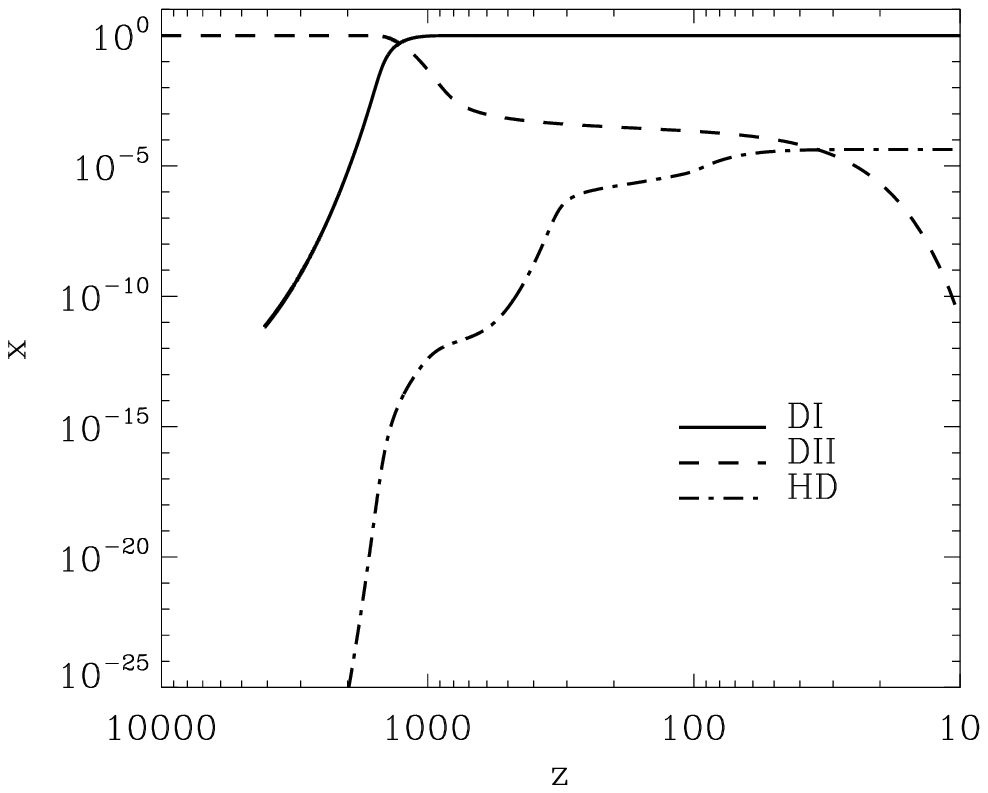}
\includegraphics[width=0.32\textwidth]{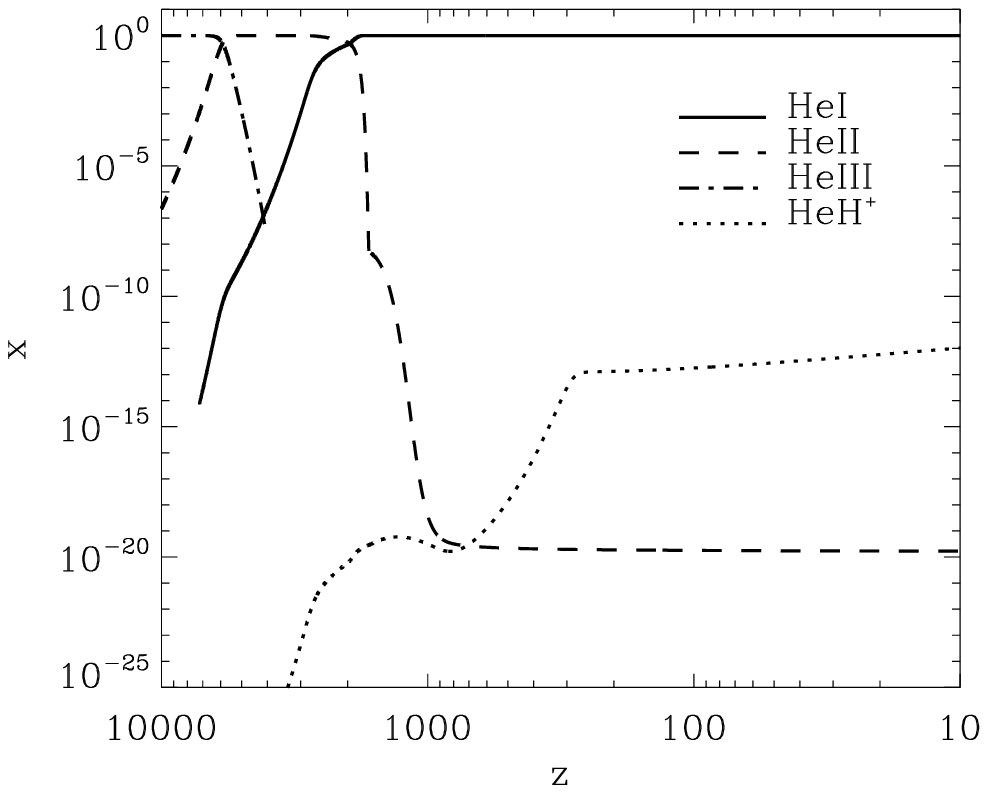}
\caption{Evolution of abundances for Hydrogen, Deuterium and Helium species.}
\end{figure}

In Fig. \ref{HDHe} the evolution of abundances $x_{HII}$, 
$x_{DII}$, $x_{HeII}$, $x_{H^-}$, $x_{H_2^+}$, $x_{H_2}$, $x_{HD}$
$x_{HeH^+}$ is shown for computation made in the framework of full modified model of the effective 3-level atom for the $\Lambda$CDM model with the best-fit parameters obtained from the \textit{Planck} satellite data (year 2015) \cite{planck2015} on CMB temperature fluctuations and polarization at all multipoles (\textit{Planck} TT,TE,EE+lowP): physical densities of baryons and cold dark matter $\Omega_bh^2=0.02225\pm0.00016$ and $\Omega_ch^2=0.1198\pm0.0015$, cosmological constant energy density $\Omega_{\Lambda}=0.6844\pm0.0091$, dimensionless Hubble constant $h=0.6727\pm0.0066$ ($\pm\,1\sigma$ confidence ranges). 
The best-fit fractions of Deuterium and Helium are $Y_p\equiv 4n_{He}/n_b=0.24667\pm0.00014$ 
($n_b$ -- baryons number density) and $f_D=(2.614^{+0.057}_{-0.060})\times10^{-5}$ 
(for the case of standard Big Bang Nucleosynthesis). Here and below the temperature of CMB is $T_{cmb}=2.7255\,K$. The values of abundances at different redshifts after the cosmological recombination are presented in the Table \ref{h2}.

We see that the abundances of $HII$, $DII$, $HeII$ and molecules, negative Hydrogen ions and molecular ions at the redshifts $z=1000$, 100 and 10 exceed the corresponding values in \cite{vonlanthen2009} where the simplified description of cosmological recombination of Hydrogen, Deuterium and Helium was used.

\begin{table}[ht]
\caption{Abundances of atoms, ions and molecules at the redshifts 
$z=1000$, $z=100$ and $z=10$}
\label{h2}
\centering
\begin{tabular}{l l l l l }

\hline
Species && $z=1000$ & $z=100$ & $z=10$ \\
\hline 
${\rm HI}$ && $0.9511$ & $0.9997$ & $0.9998$  \\
${\rm HII}$ && $4.893\times10^{-2}$ & $2.748\times10^{-4}$ & 
$2.003\times10^{-4}$  \\
${\rm H^-}$ && $1.778\times10^{-19}$ & $1.630\times10^{-11}$ & 
$3.076\times10^{-13}$  \\
${\rm H_2}$ && $1.621\times10^{-13}$ & $5.940\times10^{-7}$ & 
$2.370\times10^{-6}$  \\
${\rm H_2^+}$ && $2.125\times10^{-18}$ & $1.210\times10^{-13}$ & 
$1.260\times10^{-13}$ \\
\hline
${\rm DI}$ && $0.9518$ & $0.9998$ & $1.000$ \\
${\rm DII}$ && $4.821\times10^{-2}$ & $2.127\times10^{-4}$ & 
$1.405\times10^{-11}$  \\
${\rm HD}$ && $3.975\times10^{-13}$ & $6.343\times10^{-6}$ & 
$4.268\times10^{-5}$ \\
\hline
${\rm HeI}$ && $1.000$ & $1.000$ & $1.000$  \\
${\rm HeII}$ && $3.854\times10^{-19}$ & $1.796\times10^{-20}$ & 
$1.678\times10^{-20}$ \\
${\rm HeH^+}$ && $3.138\times10^{-20}$ & $1.772\times10^{-13}$ & 
$1.036\times10^{-12}$ \\
\hline

\end{tabular}
\end{table}

Let us estimate the impact of accuracy of computation of the cosmological recombination on number densities of the primordial molecules. To do this we compare the values obtained in the framework of basic modified model of the effective 3-level atom for both Hydrogen and Helium with the corresponding values obtained in the framework of full modified model of the effective 3-level atom. If we use the full model for Hydrogen, Deuterium and Helium, the abundances differ from the obtained in the framework of basic model ones by no more than 2\% for $H^-$, $H_2^+$, $H_2$, $HeH^+$ at redshifts from 1000 to 10  and by no more than 3\% for the molecule $HD$ at redshifts from 100 to 10. When the full model is used only for Hydrogen and Deuterium and the basic one is used for Helium, the deviations of abundances do not exceed, as in previous case, 2\% for $H^-$, $H_2^+$, $H_2$, $HeH^+$ and 3\% for $HD$. If, conversely, the recombination of Helium is described in the framework of full model while the recombination of Hydrogen and Deuterium in the framework of basic model, the deviations are negligibly small for all molecules, negative Hydrogen ions and molecular ions at redshifts from 1000 to 10.

Let us now investigate the effect of values of cosmological parameters on the abundances of $H^-$, $H_2^+$, $H_2$, $HD$ and $HeH^+$ after the cosmological recombination. Consider the $\Lambda$CDM models with the best-fit parameters obtained from: (1) \textit{Planck} satellite data (year 2015) on CMB temperature fluctuations at all multipoles and polarization at low multipoles (\textit{Planck} TT+lowP) \cite{planck2015}: $\Omega_bh^2=0.02222$, $\Omega_ch^2=0.1197$, $\Omega_{\Lambda}=0.685$, $h=0.6731$; (2) \textit{Planck} satellite data (year 2013) on CMB temperature fluctuations at all multipoles and WMAP satellite 9-year data on polarization at low multipoles (\textit{Planck}+WP) \cite{planck2013}: $\Omega_bh^2=0.02205\pm0.00028$, $\Omega_ch^2=0.1199\pm0.0027$, $\Omega_{\Lambda}=0.685^{+0.018}_{-0.016}$, $h=0.673\pm0.012$ (1$\sigma$ uncertainties). Deviations of the abundances from the corresponding values in model with the best-fit parameters \textit{Planck} TT,TE,EE+lowP for $H^-$, 
$H_2^+$, $H_2$ are up to 0.4\% in the case of model with the parameters \textit{Planck} TT+lowP and up to 1.5\% in the case of model with the parameters \textit{Planck}+WP, for $HeH^+$ they are up to 0.3\% in the case of model with the parameters \textit{Planck} TT+lowP and 1\% in the case of model with the parameters \textit{Planck}+WP at redshifts from 1000 to 10, for $HD$ they do not exceed 0.4\% at $z>300$, 0.125\% at $z>90$ in the case of model with the parameters \textit{Planck} TT+lowP and 1.5\% at $z>300$, 0.4\% at $z>60$ in the case of model with the parameters \textit{Planck}+WP and are negligibly small at lower redshifts. For models with the values of parameters at either upper or lower limits of 1$\sigma$ confidence ranges from the \textit{Planck} TT,TE,EE+lowP data the deviations of abundances from values in model with the best-fit parameters at $z<1000$ are up to 2\% for $H^-$, $H_2^+$ and $HeH^+$ and 0.6\% for 
$H_2$ and $HD$.

The values of Deuterium and Helium fractions affect the abundances of negative Hydrogen ions, molecular ions and molecules as follows. Deviations of the $x$ values in model with the best-fit parameters \textit{Planck} TT,TE,EE+lowP, best-fit Deuterium fraction and Helium fraction at either upper or lower limit of 1$\sigma$ confidence range ($Y_p=0.24743$ or $Y_p=0.24591$ 
\cite{planck2015}) from the corresponding values in model with the same set of parameters and best-fit Helium fraction are up to $\sim0.5\%$ for $H_2^+$ at $z<100$ and not more than 0.2\% for $H^-$, $H_2$, $HD$, $HeH^+$ at $z<1000$. In the case of Deuterium fraction at either upper or lower limit of 1$\sigma$ confidence range ($f_D=2.801\times10^{-5}$ or $f_D=2.424\times10^{-5}$ \cite{planck2015}) and best-fit Helium fraction the corresponding deviations are negligible for all molecules, negative Hydrogen ions and molecular ions from the cosmological recombination epoch to redshift 10. For the models with both Deuterium and Helium fractions at either upper or lower limits of 1$\sigma$ confidence ranges the deviations are up to $\sim0.5\%$ for $H_2^+$ at redshifts lower than 100 and do not exceed 0.2\% for other molecules, negative Hydrogen ions and molecular ions at redshifts lower than 1000. Deviations of the abundances in models with Deuterium and Helium fractions as well as cosmological parameters at either upper or lower limits of 1$\sigma$ confidence ranges from the corresponding values in models with all best-fit parameters are up to 2\% for $H^-$, $H_2^+$, $HeH^+$ and do not exceed 0.6\% for $H_2$ and $HD$ at all $z$ from 1000 to 10.

All previous computations have been done for the number of neutrino species $N_{eff}=3$. However, from the \textit{Planck} TT,TE,EE+lowP data it follows that $N_{eff}=2.99\pm0.20$. In case of the best-fit values of the cosmological parameters and Deuterium and Helium fractions the effect of deviation of the best-fit number of neutrino species from 3 on the abundances does not exceed 0.1\% for all molecules, molecular ions and negative Hydrogen ions. In case of the number of neutrino species at either upper or lower limit of 1$\sigma$ confidence range the corresponding deviations of abundances are up to 0.3\%. The deviations of abundances in the models with number of neutrino species, Deuterium and Helium fractions and cosmological parameters at either upper or lower limits of 1$\sigma$ confidence ranges from the corresponding values in the model with best-fit cosmological parameters, Deuterium and Helium fractions and $N_{eff}=3$ do not exceed 2.3\% for $H^-$, $H_2^+$ and $HeH^+$ and 0.6\% for $H_2$ and $HD$ at all redshifts from 1000 to 10.

Let us consider now the models with non-standard number of neutrino species. In the case of $N_{eff}=4$ (1 additional neutrino species) the deviations of abundances from the corresponding values in the model with best-fit cosmological parameters and fractions of Deuterium and Helium do not exceed  1.2\% for $H^-$, $H_2^+$ and $HeH^+$, 1.1\% for $HD$ and 0.65\% for $H_2$ at $z$ from 1000 to 10. In the case of $N_{eff}=5$ (2 additional neutrino species) the deviations of abundances do not exceed 2.4\% for $H^-$ and $H_2+$, 2.3\% for $HeH^+$, 2.2\% for $HD$ and 1.2\% for $H_2$, while in the case of $N_{eff}=6$ (3 additional neutrino species) -- 3.5\% for $H^-$ and $H_2+$, 3.4\% for $HeH^+$, 3.2\% for $HD$ and 1.8\% for $H_2$ correspondingly.

Let us discuss now how the number densities of first molecules can be affected by taking into account the chemical reactions that are not included in the minimal model. For example along with the reactions (H5), (H10), (H15), (D8) and (D10) defining the number densities of molecules $H_2$ and $HD$ in the minimal model let us consider the following reactions:
\begin{itemize}
 \item (AH1) $H_2+H\rightarrow H+H+H$ (reaction rate from \cite{coppola2011}),
 \item (AH2) $H_2+\gamma\rightarrow H+H$ (direct photodissociation, reaction rate from \cite{coppola2010} for Lyman system),
 \item (AH3) $H_2+\gamma\rightarrow H+H$ (direct photodissociation, reaction rate from \cite{coppola2010} for Werner system),
 \item (AH4) $H_2+\gamma\rightarrow H+H$ (indirect photodissociation - Solomon process,reaction  rate from \cite{coppola2011}),
 \item (AD1) $HD+H\rightarrow H+H+D$ (reaction rate from \cite{coppola2011} -- similarly to \cite{gloabel08} we use for this process the reaction rate (AH1)),
 \item (AD2) $HD+\gamma\rightarrow H+D$ (direct photodissociation, reaction rate from \cite{coppola2010} for Lyman system),
 \item (AD3) $HD+\gamma\rightarrow H+D$ (direct photodissociation, reaction rate from \cite{coppola2010} for Werner system).
\end{itemize}
The contributions of these reactions to the changes of number densities of $H_2$ and $HD$ are presented in the Table \ref{react}. We see that the contributions of reactions (H10) and (H15) exceed the contributions of reactions (AH1)-(AH4) while the contributions of reactions (D8) and (D10) -- the contributions of reactions (AD1)-(AD3) correspondingly at all redshifts from 1000 to 500. However, the reactions (H10) and (H15) are opposite as well as the reactions (D8) and (D10), so the rate of growth of the number densities of $H_2$ and $HD$ is determined by the difference (H10)-(H15) and (D8)-(D10) respectively (for $H_2$ the growth of number density is also sped up by the reactions (H5) and (D10) and slowed down by the reaction (D8)). Due to the low temperature of CMB the contribution of photodissociation of $H_2$ at $z=1000$ does not exceed 0.25\% of the sum of contributions of reactions (H5)+(H10)-(H15)-(D8)+(D10) and the contribution of photodissociation of $HD$ does not exceed 0.2\% of the difference of contributions of reactions (D8)-(D10). The contributions of photodissociation decrease with decreasing of $z$. Thus, the photodissociation of molecules $H_2$ and $HD$ can be neglected at the redshifts less than 1000. At $z=1000$ the contribution of reaction (AH1) makes $\sim55\%$ of the sum of contributions (H5)+(H10)-(H15)-(D8)+(D10), while the contribution of reaction (AD1) $\sim55\%$ of the difference of contributions (D8)-(D10). So, at the redshift 1000 taking into account the reactions (AH1) and (AD1) will slow down the formation of molecules $H_2$ and $HD$ approximately by factor of 2. At $z=900$ the contributions of reactions (AH1) and (AD1) do not exceed $15\%$ and at $z=800$ $2\%$ of the sum of contributions (H5)+(H10)-(H15)-(D8)+(D10) and of the difference of contributions (D8)-(D10) respectively. Already at $z=700$ the contributions of reactions (AH1) and (AD1) are negligible.

\begin{table}[ht]
\caption{Contributions of the chemical reactions ($s^{-1}$) from the Table \ref{reaction} and the additional reactions not included in the minimal model to the redshift change of number densities of the molecules $H_2$ and $HD$. For collisional processes: $k_{ij}x_ix_j$, for radiation ones: $k_{i\gamma}x_i$. Contributions of the reactions for $HD$ are multiplied by $f_D$.}
\label{react}
\centering
\footnotesize
\begin{tabular}{l l l l l l l}

\hline
Reactions&$z=1000$&$z=900$&$z=800$&$z=700$&$z=600$&$z=500$ \\
\hline 
(H5)&$3.3283\times10^{-26}$&$8.3908\times10^{-27}$&$2.2116\times10^{-27}
$&$9.5927\times10^{-28}$&$6.3109\times10^{-28}$&$5.7183\times10^{-28}$\\
(H10)&$2.4427\times10^{-25}$&$1.1466\times10^{-25}$&$6.8675\times10^{-26}
$&$9.0506\times10^{-26}$&$2.8233\times10^{-25}$&$2.5080\times10^{-24}$\\
(H15)&$2.0023\times10^{-25}$&$3.8589\times10^{-26}$&$4.3035\times10^{-27}
$&$4.9211\times10^{-28}$&$6.1995\times10^{-29}$&$1.0471\times10^{-29}$\\
\hline
(D8)&$8.0983\times10^{-26}$&$3.6709\times10^{-26}$&$1.1987\times10^{-26}
$&$5.5183\times10^{-27}$&$4.5474\times10^{-27}$&$1.1214\times10^{-26}$\\
(D10)&$8.0978\times10^{-26}$&$3.6704\times10^{-26}$&$1.1982\times10^{-26}
$&$5.5117\times10^{-27}$&$4.5267\times10^{-27}$&$1.1025\times10^{-26}$\\
\hline
(AH1)&$4.2816\times10^{-26}$&$1.1225\times10^{-26}$&$1.1609\times10^{-27}
$&$5.0826\times10^{-29}$&$9.3180\times10^{-31}$&$6.8023\times10^{-33}$\\
(AH2)&$1.5708\times10^{-30}$&$5.6162\times10^{-33}$&$2.7681\times10^{-36}
$&$1.2587\times10^{-40}$&$2.4497\times10^{-46}$&$5.2153\times10^{-54}$\\
(AH3)&$4.6770\times10^{-31}$&$1.3254\times10^{-33}$&$4.8805\times10^{-37}
$&$1.5233\times10^{-41}$&$1.7918\times10^{-47}$&$1.8805\times10^{-55}$\\
(AH4)&$1.7650\times10^{-28}$&$1.2925\times10^{-30}$&$1.5086\times10^{-33}
$&$1.9160\times10^{-37}$&$1.2631\times10^{-42}$&$1.0332\times10^{-49}$\\
\hline
(AD1)&$2.7444\times10^{-30}$&$7.3146\times10^{-31}$&$7.7265\times10^{-32}
$&$3.4757\times10^{-33}$&$6.5943\times10^{-35}$&$5.0173\times10^{-37}$\\
(AD2)&$8.3736\times10^{-33}$&$5.3323\times10^{-35}$&$5.4243\times10^{-38}
$&$6.2807\times10^{-42}$&$4.2604\times10^{-47}$&$5.2039\times10^{-54}$\\
(AD3)&$5.2459\times10^{-34}$&$2.1056\times10^{-36}$&$1.1984\times10^{-39}
$&$6.5459\times10^{-44}$&$1.6204\times10^{-49}$&$4.7849\times10^{-57}$\\
\hline

\end{tabular}
\end{table}

\section*{Conclusions}

Calculations of abundances of the first molecules based on modified model of the effective 3-level atom for Hydrogen, Deuterium and Helium for cosmological recombination and the minimal model for kinetics of chemical reactions in the Dark Ages epoch have shown that in the early Universe the fractions of formed $H_2$ molecules was the largest. At $z=10$ the abundance of $H_2$ exceeds the abundances of $H_2^+$ and $HeH^+$ by $10^7$ times, $H^-$ by $10^6$ times, $HD$ by $10^3$ times.

Inaccuracies in the computation of cosmological recombination lead to uncertainties of the abundances of molecules, negative Hydrogen ions and molecular ions not exceeding 2-3\%, but comparable to the uncertainties caused by the uncertainties of values of cosmological parameters (up to 2\%). The accurate description of Hydrogen and Deuterium recombination is the most important thing. The effect of uncertainties in determination of the number of neutrino species as well as of the Deuterium and Helium fractions from the \textit{Planck} satellite data are marginal (uncertainties of abundances $\lesssim0.5\%$).

After the end of Dark Ages epoch and the beginning of Cosmic Dawn ($z<30$) it is necessary to take into account the reionization of the medium, but the ambiguities in theoretical description of this process do not allow to obtain sufficiently accurate number densities of ions and primordial molecules.

\appendix

\section{Number densities of molecules during the equilibrium recombination}\label{a}

At the stage of equilibrium recombination of Hydrogen, Deuterium and Helium from the equations
(\ref{cin}) and the condition
\begin{eqnarray*}
\frac{dx_i}{dt}=0
\end{eqnarray*}
the system of algebraic equations for the abundances of molecules, negative Hydrogen ions and molecular ions can be obtained. It can be linearized assuming that $x_{HII}$, $x_{DII}$ and $x_{HeII}$ are zero-order quantities and 
$x_{H^-}$, $x_{H^+_2}$, $x_{H_2}$, $x_{HD}$, $x_{HeH^+}$ are first-order quantities. The solutions of the linear system of equations are as follows:
\begin{eqnarray*}
&&x_{HeH^+}=\frac{N^0_{HeH^+}}{D_{HeH^+}},\\
&&N^0_{HeH^+}=k_{HeIHII}n_Hx_{HII}\left(1-x_{HeII}\right),\\
&&D_{HeH^+}=k_{HeIHII}n_Hx_{HII}+k_{HeH^+\gamma}+k_{HeH^+HI}n_H\left(1-x_{HII}
\right),\\
&&x_{H^-}=\frac{N^0_{H^-}}{D_{H^-}},\\
&&N^0_{H^-}=k_{HIe}\left(1-x_{XII}\right)\left(x_{HII}+x_{DII}f_d+x_{HeII}f_{He}
\right)n_H,\\
&&D_{H^-}=k_{H^-\gamma}+k_{H^-HI}\left(1-x_{HII}\right)n_H+k_{H^-HII}x_{HII}n_H,
\\
&&x_{H^+_2}=\frac{N^0_{H^+_2}+N^{H^-}_{H^+_2}x_{H^-}}{D_{H^+_2}},\\
&&x_{H_2}=\frac{N^{H^-}_{H_2}D_{N^+_2}+N^{H^+_2}_{H_2}N^{H^-}_{H^+_2}}{D_{H_2}D_
{H^+_2}}x_{H^-}+\frac{N^{H^+_2}N^0_{H^+_2}}{D_{H_2}D_{H^+_2}},\\
&&N^0_{H^+_2}=k_{HIHII}x_{HII}\left(1-x_{HII}\right)n_H+\left(k_{HeH^+HI}
\left(1-x_{HII}\right)-k_{HIHII}x_{HII}\right)\\
&&\times f_{He}n_HN^0_{HeH^+}/D_{HeH^+},\\
&&N^{H^-}_{H^+_2}=\left(\left(k_{H_2HII}-2k_{HIHII}\right)x_{HII}-k_{HIHII}k_{
DIIH_2}x_{DII}f_D/k_{HDHII}\right)n_HN^{H^-}_{H_2}/D_{H_2}\\
&&-k_{HIHII}x_{HII}n_H,\\
&&D_{H^+_2}=2k_{HIHII}x_{HII}n_H+k_{H^+_2\gamma}+k_{H^+_2HI}\left(1-x_{HII}
\right)n_H\\
&&-\left(\left(k_{H_2HII}-2k_{HIHII}\right)x_{HII}-k_{HIHII}k_{DIIH_2}x_{DII}
f_D/k_{HDHII}\right)n_HN^{H^+_2}_{H_2}/D_{H_2},\\
&&N^{H^-}_{H_2}=k_{H^-HI}\left(1-x_{HII}\right),\\
&&N^{H_2^+}_{H_2}=k_{H_2^+HI}\left(1-x_{HII}\right),\\
&&D_{H_2}=k_{H_2HII}x_{HII},\\
&&x_{HD}=\frac{k_{DIIH_2}x_{DII}}{k_{HDHII}x_{HII}}x_{H_2}.\\
\end{eqnarray*}

\section*{Acknowledgements}
This work was supported by the project of Ministry of Education and Science of Ukraine (state registration number 0116U001544).

\end{document}